# A New Global Theory of the Earth's Dynamics :

## a Single Cause Can Explain All the Geophysical and Geological Phenomena

André Rousseau

*CNRS (UMS 2567) Université Bordeaux 1, Groupe d'Etude des Ondes en Géosciences*

*351 cours de la Libération, F-33405 Talence cedex*

*a.rousseau@geog.u-bordeaux1.fr*

After describing all the contradictions associated with the current Plate Tectonics theory, this paper proposes a model where a single cause can explain all geophysical and geological phenomena. The source of the Earth's activity lies in the difference of the angular velocities of the mantle and of the solid inner core. The friction between both spheres infers heat, which is the cause of the melted iron which constitutes most of the liquid outer core, as well as the source of the global heat flow. The solid inner core angular velocity is supposed to remain steady, while the mantle angular velocity depends on gyroscopic forces (involving acceleration) and slowing down due to external attractions and, principally the motions of mantle plates 2900 km thick.

The variations of the geomagnetic field are therefore the direct consequence of the variations of the angular velocity of the mantle relative to that of the inner core. As a result, the biological and tectonic evolutions during geological times are due to those phenomena. So, the limits of eras coincide exactly with the passage to zero of the geomagnetic field.

Here we show that cycles of about 230-250 millions years, which exhibit the correlation between the mantle angular velocity variations, the geomagnetic variations, and therefore the climate, allow us to predict future events : the current global warming which parallels the Earth's magnetic field decrease, and the spoliation of the forests which are not damaged by acid rains, but in reality from their roots by acid liquid arising inside the cracks of the crust.

**Finally, this geodynamics allows us to determine the mechanisms of earthquakes.**

**Keywords**

Geodynamics, geomagnetism, orogeneses, core, mantle, thermoelectricity

Since the discovery of the mid-oceanic ridges in the 1960's the continental drift model, proposed by Wegener in the 1930's and rejected at that time by the geoscientists, has been transformed into "Global Plate Tectonics" with the mechanical support of convection currents inside the mantle. But the continuing impossibility of predicting earthquakes reveals an insufficient knowledge of Earth's geodynamics, which can be compared to the state of knowledge in medicine before Pasteur's discovery of micro-organisms. As a confirmation of this comparison, Plate Tectonics must be continuously modified in order to "explain" an earthquake occurring in an unexpected area (such as the last Kobe earthquake, for example). This reminds one of the medieval period when the anti-Copernic astronomers had to invent the concept of epiglyptical movements of the planets in order to adjust the observations to the geocentric model of the Universe.

That is why it would be wise to consider inadequate the empirical approach to the lithospheric plates defined by some statistical outlines of earthquake epicentres. Besides, in spite of the public success of Plate Tectonics theory, some authors, as A.A. Meyerhoff, pointed out the contradictions with reality and proposed a new global geodynamics, the Surge Tectonics (see *Meyerhoff et al. (1996)*), based on magmatic activity through the lithosphere, from the asthenosphere inside the upper mantle. Consequently, attempting to invent a possible mechanical model of the Earth's geodynamics, after abstracting from observed effects, may be fruitful if the consequences of this model fit the observed reality. If only one cause can explain the majority of the observations, which this paper intends to demonstrate, the probability that the theory is grounded in reality is great.

A.A. Meyerhoff located the "engine" of the Earth's dynamics at the bottom of the upper mantle. The model developed here "puts" it deeper, in the liquid outer core, which may be seen as a kind of magma. This model is based on the hypothesis of differences in the angular velocities of the solid inner core and of the mantle for mechanical reasons. This difference has already tentatively been put forward by several authors *(Song and Richards, 1996 ; Su et al., 1996 ; Glatzmaier and Roberts, 1996 ; Jeanloz and Romanowicz, 1997 ; Aurnou et al., 1998)*. We intend to demonstrate that, in fact, such a phenomenon is essential. The heat emitted by the resulting friction causes the melting of the outer core, geothermal effect, and hot spots. The differences between the tangential velocities at the equator and at the poles of the surface of the inner core and of the periphery of the outer core infer a maximum friction at the equator tending to zero at the poles, and as a result producing a decreasing heat gradient. Consequently two electric fields originate for thermo electrical reasons from these two shells, which may be the source of the geomagnetism and explain its variations. However, magneto-hydro dynamics effect could also be a source of the geomagnetism under specific conditions.

The concept of lithospheric plates 100 km thick is replaced by mantle plates 2900 km thick. From the variations of the geomagnetic field due to the variations in the angular velocity of the mantle, we show the consequences on climate and species during the geological times. The main orogeneses and glaciations periods are connected to the Earth's magnetism and prove to be integrated in a regular time cycle. Moreover earthquakes are not only caused by a plate confrontation.

We present a calculation related to the magnetic field of the model, which makes it plausible. Assuming steady the present decrease of the Earth's angular velocity, which we only attribute to the mantle, we calculate the probable constant inner core angular velocity, and the electric charges responsible of the Earth's magnetism.

**Major problems of the current Plate Tectonics theory**

Following are the principle failings of the current Plate Tectonics theory.
1) The basic concept which supports the current Plate Tectonics theory is the assumption of convection movements located inside the mantle and caused by the heat of the core. There is however a contradiction in the fact that the mantle lets shear waves propagate inside it, which reveal a crystalline structure inconsistent with the possibility of convection motions inside fluids.

2) The oceanic bottoms are flat and the sediments are not folded. We should take into account the pressure and constraints that the up-lifting and superficial parts of those convection currents would apply on the thin lithospheric Plates which are 100 kilometres thick and several thousands kilometres long ; it is not conceivable that these plates can exist without being at least crumpled.

3) The Benioff or "subduction" planes along which there would be the lithospheric slabs entering the mantle are said to be the location of earthquakes and active volcanoes : the evoked cause would be the mechanical effect and the heating due to the friction created by the downward slab. But there is inconsistency between mechanical ruptures (such as earthquakes) and melted rocks (such as magmas). Besides, the supposed mantle local magmatic zones ought to be cooled down by the cold elements introduced by downward slabs, which come from the surface, unless there are already melted zones extending round about at the depth of 100 kilometres. But it is unlikely, otherwise this would have been revealed by the lack of shear waves - nevertheless present -, as in the case of the fluid characteristics of the outer core. In addition, such a model does not explain the characteristic shape of island arcs.

4) The geothermal flux is reputed to be same all over the globe, and to be the consequence of the radioactive activity of the uranium present in crystalline rocks ; however, those rocks do not exist in the oceans (2/3 of the Earth's surface).

5) The inner core is solid and the outer core liquid ; as both parts are reputed to have the same composition, the opposite would be more logical given a central heat source. If the very high pressure was the cause of this state of fact, we could not have the existing sharp seismic interface between the outer and inner core.

6) One of the most upsetting enigmas is what occurs when the Earth's magnetic field is equal to zero, particularly in relation to the magnetosphere and the solar flux. One cannot be content with estimating this period too short to involve some consequences linked to the quantity of the solar flux hitting then the Earth's surface. The number of magnetic inversions should be examined primaruly from the volcanic samples, and not from the oceanic surface "inversions", the origin of which is not assured.

The oceanic bottoms are indeed more complicated than the descriptions made in accordance with the "canons" of Plate tectonics. So, Cretaceous sediments (instead of Miocene or later) and even sericitic carbonaceous phyllites and microquartzites, as well as carbonaceous shales were dredged in the equatorial segment of the mid-Atlantic ridge, where the investigation of the igneous rocks reveals a continental crustal relict of pre-Cretaceous or pre-Jurassic age *(Udintsev et al., 1992 ; Udintsev and Kurentsova, 1993)*. Thus, it is difficult to attribute the variations of the oceanic magnetic field to the differences in the paleomagnetic directions of oceanic bed rock ; they are more probably due to the contact between paramagnetic peridotites and magnetic serpentinites (altered peridotites).

7) Finally, the cause of earthquakes occurring inside Plates which are supposedly impossible to distort should not be that of rubbing Plates.

**A new Geodynamic Model for the Earth**

Let us *imagine* the evolution of a very young spherical terrestrial planet of 6370 kilometres in radius, moved by its initial angular movement.

• First step : The gravitational differentiation of its constituents causes the heaviest of them (density about 7 for example) to be moved and concentrated in the central area of the sphere.

• Second step : There is a sharp gravitational separation between this heavy central area roughly 3500 kilometres in radius and the envelope of lower density (3.3 for example). The external attractions acting on the Earth's envelope (the moon and near planets), and the differences in the respective kinetic energies and the centrifugal forces of both masses in rotation, involve contradictory constraints at the limit between them. The result is a curbing of the envelope rotational velocity, and, as a consequence, a fragile zone arises at the interface.

• Third step : This fragile interface breaks, which makes both parts independent from each other. Let us call them the core inside and the mantle outside. The angular velocity of the core is supposed to remain constant, as well as its rotational axis which represents the inertial movement from the solar system at the moment of its creation. Then, the mantle, more than 50 % lighter and thinner (2900 kilometres compared to 3500) can slow down more easily than before.

• Fourth step : The difference in the angular velocities of both masses involves friction, which increases the heat which reaches the lowest fusion point of both constituents. Therefore, if the core is made of iron and the mantle of aluminosilicates, the iron will tend to fuse first, because its fusion point is lower at equal pressure. We will then have an outer liquid core and an inner solid core.

• Fifth step : This liquid helps to lubricate the zone located between the mantle and the solid core – which tends to externally melt more and more. The result is an acceleration of the difference between the angular velocity of the mantle and that of the solid core. Under gyroscopic forces the mantle rotates faster than the solid core, as the general trend of the air masses within the troposphere is a displacement from West to East more rapid than the Earth's rotation which is responsible for this effect. But the Coriolis forces are different according to the latitudes, as well as the external attractions of the moon and near planets. Therefore, the mantle undergoes contradictory forces, which leads to cracks and breaks ; a few independent plates 2900 kilometres thick result, and begin to drift above the liquid core, essentially moved by the Coriolis forces. Between those plates, the liquid of the outer core takes advantage of the gaps in order to reach the surface, assimilating on the way the material of the mantle.

• Sixth step : As the mantle plates float and move more and more rapidly, they absorb more and more energy, the result of which is the slowness of the angular velocity of the mantle, which becomes slower than that of the solid core.

• Seventh step : The plates begin to collide, then their motions are stopped. If almost all the plate motions are stopped, one of the brakes on the mantle rotation disappears, and then this rotation begins to accelerate again, until the motions of the new plates brake the mantle rotation. A new cycle begins.

The peripheral melting of the solid core must modify its radius in proportion of the friction heat quantity, that is to say of the differences in the angular velocities of the core and the mantle. Due to energy conservation, the solid core angular velocity should vary at least a little in respect to its volume since its density remains steady. But, in this paper, this phenomenon will not be taken into account in order to avoid introducing too many unknown quantitative parameters.

**Possible internal consequences of such a model**

This model can involve some specific consequences in several fields of the Earth Sciences.

• Because the liquid outer core is the seat of the heat source generated by the friction between the inner core and the mantle, the temperature should decrease towards the centre of the solid inner core, and more rapidly does towards the Earth's surface through the mantle. Thus it is logical that the inner core is solid and the outer core liquid.

• The hot melted material of the outer core tends to rise between the edges of the new plates and then assimilates on the way some mantle material. After arriving at the Earth's surface, this magma accumulates and creates mountains - such as oceanic ridges - accompanied by their corresponding rifts, which are the vertical channels of the up-going material.

• The heat flow towards the surface will be the largest above the up-going magmas and steady between them. In other words, the geothermal flow must be globally constant and the same above continents as well as above oceans.

• The source of the Earth's magnetism and of the telluric currents may be then contemplated from the conductive masses in movement. The friction between the masses of different angular velocities is maximum at the equators and tends to zero toward the poles and so does the consecutive heating. Thus, thermo electrical gradients arise with electrical charges at the surface of the conductive solid inner core and the periphery of the conductive liquid outer core. These two shells rotating at different velocities create the geomagnetic field, the variations of which depend on the difference between both velocities (see further the calculations and discussion). As it is assumed that the inner core velocity remains probably steady in time, the variations of the mantle angular velocity are responsible of the geomagnetic field variations, particularly of the magnetic reversals. When the mantle and the core rotate at the same velocity, the geomagnetic field is equal to zero.

- The rotational axes of the inner core and of the mantle may be different. Particularly the one of the mantle may be variable ; this mantle, indeed, is more sensitive than the solid inner core to the various external attractions coming from the Solar system, because it "floats" above a liquid. At the Earth's surface an apparent polar nutation ensues, and the geomagnetic poles may wander, reflecting the instability of the electrically charged shell located at the basis of the mantle, probably the origin of the dipolar field (see further), the non dipole field being probably consecutive to the electrically charged shell located at the surface of the inner core.

- As a consequence of the existence of a magnetic field, electrical (so-called telluric) currents propagate in a direction perpendicular to the magnetic field and eastwards in the conductive areas of the mantle and the Crust, particularly inside fluids. There must be a radial decrease of the electric potentials towards the surface, which must cause an electric current when two conductive horizontal layers located at different levels are connected by a conductor. In this case, the spontaneous potential often observed along a slope can be easily explained, as well as in vertical boreholes. At the Earth's surface, the pointed shapes accumulate electrical charges like capacitors, which may involve sparks (so-called lightning) with aerial opposite charges. Theoretically, the vertical cliffs directed eastwards ought to accumulate electrical charges too, and it is easy to observe that large radio waves are more disturbed above the cliffs directed eastwards than above those directed towards other directions.

- The plates as they have been defined above, that is to say 2900 km thick, may be fissured vertically. The hot liquid material of the outer core may then penetrate into the mantle basis, and continue its penetration up to the Crust, by melting and assimilating the mantle material while moving forward, which causes some areas called "hot spots" from the surface (as the Yellowstone geysers, for example). Various kinds of chemical composition can be transported by gas or liquid through the cracks and fissures, and reach the continental surface, even far from hot spots (see the case of the borehole KTB in Bavaria, for example), or deep oceanic troughs (as in Japan for example). That is why the phenomenon of "acid forests" may result, among other causes, from this kind of acid liquid oozing up to tree roots under the surface.

**Relation of this model to the geological history of the Earth**

The diagram of Fig. 1 shows one curve which represents two phenomena as a function of geological time between the hercynian orogenesis and the present time : 1) the mantle angular velocity variations, and 2) the magnitude and directions of the geomagnetic field.

The major orogeneses are the final result of the mantle plates defined above colliding with themselves. Their horizontal motions are then stopped. When this occurs, the motions of these mantle plates have slowed down the mantle angular velocity which has become slower than the solid core, and has consequently caused the reversal of the geomagnetic field. Afterwards, as soon as the orogenesis is "consolidated", after creating new plates which do not yet move at this time, the angular velocity of the mantle must increase rapidly, and become faster than that of the solid core. Correlatively, the absolute value of the (reverse) geomagnetic field decreases down to zero when the angular velocities of the mantle and of the solid core are almost equal. Afterwards the geomagnetic field increases northward until the mantle rotation is sufficiently slowed by the motions of the plates to begin a slow deceleration. The corresponding geomagnetic field then decreases and reverses after passing zero. Afterwards, the cycle begins again.

When the geomagnetic field is large, there is a maximal deviation of the solar flux towards the poles by the magnetosphere, decreasing the heat at the Earth's surface. This involves global cooling and large glaciations, which occurred in the early quaternary.The Triassic glaciations left some indications such as tillites (at Granville in France, for example).

As for the periods when the geomagnetic field is very weak and equal to zero, we have the opposite situation : there is no more magnetosphere, and the solar flux directly reaches the surface of the whole globe. As a consequence, the heat increases in the whole atmosphere, and also the ionosphere (with its van Allen belts) disappears ; the structure of the atmosphere at these times is certainly different from that of today. The densities of the present troposphere and stratosphere were perhaps smaller than today ; in this case, voluminous meteorites could more easily reach the Earth's surface. Therefore, it is easy to conceive that the changes in temperature and probably hygrometry at these times are propitious for the disappearance of species and genetic mutations. These cases correspond, in fact, to each change of geological era : on one hand, the disappearance of the armoured fishes, Graptolites, Brachiopods, great Pteridophytas..., at the Palaeozoic-Mesozoic transition ; of the Dinosaurs, Ammonoidea..., at the Mesozoic-Kainozoic transition ; of the great Mammalia at the Kainozoic-Quaternary transition ; and on the other hand, the appearance of new phyla such as the Reptiles at the Palaeozoic-Mesozoic transition, the Mammalia and Angiospermae at the Mesozoic-Kainozoic transition, the Primates at the Kainozoic-Quaternary transition. In short, the era limits, which were determined empirically, represent in fact the moments when the geomagnetic field passed zero.

The increases and decreases in the mantle velocity may not be monotonous but jerky, which has corresponding repercussions for the geomagnetic field. These phenomena are particularly perceptible in two cases : 1) when the geomagnetic field is near zero, 2) when it is maximal. In the first case, the velocity jerks allow us to explain the numerous magnetic inversions occurring after the alpine orogenesis, and also in the Trias after the hercynian orogenesis, while the Cretaceous period shows a permanent direct field, and the Permian a long reversed field. The second case allows us to understand why there were several periods of glacial periods with interglacial stages in the Quaternary : the mantle rotation velocity "hesitated" during about two million years before definitively decreasing.

The mean time intervals between the great orogeneses are, in the present state of absolute dating :

- 230 Millions Years (MY) between the alpine and hercynian orogeneses,
- 220 MY between the hercynian and caledonian orogeneses,
- 100 MY (?) between the caledonian and cadomian orogeneses,
- 400 MY between the cadomian and Grenvillian orogeneses.

In fact, as the Grenvillian orogenesis is estimated at 1 000 MY BP, it is relevant to attribute a mean orogenic frequency of about 230-250 MY. This would be needed to attribute an age dating of 750 MY to the cadomian orogenesis instead of the 600 generally attributed.

**Global geological consequences inside the mantle plates**

The plates defined in this paper, as they should have resulted from the last major orogenesis, that is to say, the alpine orogenesis, should be (partly) delimited by the oceanic and continental rifts. It is not our aim to describe all of them, but only to give some characteristic examples. We discuss a partition into plates which are much larger than the lithospheric plates in the literature ; for example, the Eastern Atlantic, Africa, Europe and the western part of Asia would represent only one mantle plate. The Mediterranean region is only a relic of the Mesogea, the ophiolitic zones coming from the ancient rift which was the ancient limit between Eurasia and Africa before the alpine orogenesis ; but at present both continents are completely welded. However two individualised mantle plates may be welded only on a part of their edges : for example, the Pacific mantle plate is probably welded in Siberia to the mantle plate described above and therefore without relative motion between them at this location. On the other hand, there may be friction between two mantle plates. The motions of them are

- either horizontal such as between the Northern Pacific plate and the plate composed of America (North and South) and western Atlantic, in the North-West along the San Andreas Fault zone,

- or oblique on an inclined plane, such as the southern border of this last mantle plate with the south-eastern Pacific mantle plate, along the so-called Benioff plane at the West of Chile.

In addition, such large plates may undergo inner modifications ; at least two of them are visible at the surface.

1) The author showed more than a decade ago that the confrontation of all the apparently contradictory geophysical data of the Mediterranean region could not be explained by a present plate facing, as is usually mentioned in the literature, but more probably by a present annular complex and a lopolith in the making *(Rousseau, 1987 ; Rousseau, 1992)*. The model proposes magma rising inside a chimney of 1300 kilometres of radius centred below the Aegean sea, with a helical upward motion because of the Coriolis force. After penetrating into the lithosphere, the magma falls into the upper part of the central area of this chimney, crystallising because of the cooling, involving the collapsed structure of the Aegean region, whereas there is a bumping of the lithosphere around this zone, crossed by thin columns of magma which may reach the surface (volcanoes). Inside the lithosphere, this bumping causes tensions and constraints producing normal and reverse faults, and observation of the surface motions suggests a clockwise rotation in western Europe, which causes shear fractures. The origin of this rotation may come from the rotation of the "Aegean" magma, and perhaps also from a horizontal pressure exerted northward by Africa.

2) The great fractures inside the mantle plates can be explained by the contradictory constraints they are undergoing, either on their edges by other plates, involving for example the so-called "transform faults" below the Atlantic Ocean, or in the neighbourhood of the great rising and rotating magmatic formations of the kind seen above, as in the case of the North Anatolian Fault, which does not have the same genesis as the San Andreas Fault. The North Anatolian Fault appears to be the consequence of the hindered rotation of western Europe against a stable block, the Russian platform *(Rousseau, 1987 ; Rousseau, 1992)*.

The morphological consequence of the model described above is an island arc. It is the case of the Aegean region. It is also probable that the Indonesian arc, the Japan arc, the Caribbean arcs, for example, can be explained by this kind of model, as well as atolls.

**Relations to earthquakes**

As earthquakes come from cracks inside solid rocks in the upper part of the Earth, we must examine the different kinds of mass motions which occur inside the lithosphere (crust and upper mantle) and under this lithosphere. However, observation shows that there is not necessarily direct connection neither between earthquakes and displacements, nor pre-existing fractures. If we consider earthquakes as a simple result of a crack due to opposite constraints inside a solid, we must therefore examine the following cases.

1) If magma "lubricates" the relative motions of two mantle plates, as in the case of active rifts of the mid-Atlantic oceanic ridge, earthquakes will be rather occasional, even if their magnitude may be very high. On the other hand, if rock is against rock, as for the San Andreas Fault zone, the occurrence of earthquakes will depend on the shape of the reciprocal indentations or of the crenels relative to the plate motions, as is suggested in Fig. 2. The shear crack (mode II) which permits the release of constraints accumulated by continuous relative displacement of the plates, may occur in the neighbourhood of their edges, if there is a weak point. For example, the half-dome of the Yosemite valley in California is perhaps the result of such a phenomenon. We may also have the same causes of earthquakes in the case of the fractures occurring in the North-Anatolian Fault zone.

2) If magma pushes the lithosphere upwards, as is supposed for island arcs and atolls, the tensile forces will involve cracks of mode I (so-called extension) and of mode II (so-called compressions) as a function of the kind of the deformation of the lithosphere (see Fig. 3 and Fig. 4 as the example of the Mediterranean region).

3) If torsion forces apply somewhere inside a mantle plate, because, for example, opposite forces exerted from outside or movements are thwarted, one or several shear cracks will occur in weak points, which are not necessarily pre-existing fractures. This might be the case of Western Europe. However, if an area of large horizontal extension domain results from that, fractures and graben may occur, as the Rhenan trough (France-Germany) or Death Valley (California).

**Calculations and discussion**

We present some *simple* calculations of a possible geomagnetic field in connection with our geodynamical model. The aim is to show that this model is viable, and that the results are realistic, even if they are not exactly right.

The existence of differential axial rotation movements of conductive masses inside the Earth allows us to take into consideration two effects : the magneto-hydro-dynamical effect and the thermo electrical effect. In order that the first be a possible source of the geomagnetism, there ought to be conductive veins within the mantle near its bottom which parallel the outer core. But for lack of more precise knowledge of this environment, the calculation has been applied to the model based on the properties of thermo electricity, because we deal with the warm and conductive media of the inner and outer cores. The friction between the mantle and the solid core rotating at different velocities is supposed to be the *maximum permanent* heat source inside the globe and is therefore located in the liquid outer core.

In principle, we can consider that the possible differences in the radial temperatures of the outer core are not sufficient to cause thermo electric phenomena in this direction. But, the differences in the angular velocities of the mantle and the inner core mean that the differences in *the linear (or tangential) velocities* are not same from the equatorial zones to the polar zones of both spheres : they vary from their maximum at the equator to zero at the poles. Consequently the friction is maximum at the equator and null at the poles, which causes an increased heating from the poles to the equator inside the whole outer core. However this medium, for the reason it is liquid, should be electrically neutral because the positive and negative ions are in equal number and mixed by the rotational movement. Thus, it cannot be the source of any "self-induction dynamo".

But, owing to thermo electrical effects, the electrons can migrate towards the poles in two precise locations : (i) inside the conductive medium of the surface of the inner core, which rotates at the inner core angular velocity, and (ii) inside the conductive medium of the periphery of the liquid outer core, which is "stuck" to the bottom of the mantle and rotates at its velocity. Note that these differences in heat distribution can explain the axi-symmetric anisotropy of the inner core described in the literature from seismic travel times *(Shearer et al., 1988 ; Creager, 1992 ; Tromp, 1993 ; Song and Helmberger, 1995 ; Su and Dziewonski, 1995)*.

As long as the angular velocities of the mantle and the inner core are different, there is no tendency to heat homogeneity because (i) this heat device is open towards the mantle, and (ii) the inner core is not yet saturated by heat accumulation from the calculation of the thermal characteristic time of iron which yields a time superior to the Earth's age. As a result, there are two sources of the geomagnetic field, which can be assimilated to two electrically charged shells. This can be the cause of the existence of the "dipolar" and "non dipolar" fields. The rotational axes of the inner core and of the mantle do not mix, but they are close to each other, and we will consider they are the same for our calculations of geomagnetic field magnitude. However, if the rotational axis of the mantle were very different from that of the inner core, several dozens of degrees for example, we could have a magnetic field moving like a corkscrew, as Uranus has while rotating.

The present vertical geomagnetic field Bz at the North Pole is then given by the relation

$$Bz = Bz_1 + Bz_2$$

where $Bz_1$ is the contribution of the surface of the inner core and $Bz_2$ that of the outer core periphery.

$$Bz_1 = \frac{\mu_0}{2} R_1^4 \Omega_1 \left\{ \rho_{11} \int_0^{\alpha_1} \frac{\sin^3\theta \, d\theta}{(R_1^2 + h^2 - 2R_1 h \cos\theta)^{3/2}} + \rho_{12} \int_{\alpha_1}^{\pi-\alpha_1} \frac{\sin^3\theta \, d\theta}{(R_1^2 + h^2 - 2R_1 h \cos\theta)^{3/2}} + \rho_{11} \int_{\pi-\alpha_1}^{\pi} \frac{\sin^3\theta \, d\theta}{(R_1^2 + h^2 - 2R_1 h \cos\theta)^{3/2}} \right\}$$

where
$$\rho_{12} = -\frac{1-\cos\alpha_1}{\cos\alpha_1} \rho_{11}$$

and

$$Bz_2 = \frac{\mu_0}{2} R_2^4 \Omega_2 \left\{ \rho_{21} \int_0^{\alpha_2} \frac{\sin^3\theta \, d\theta}{(R_2^2 + h^2 - 2R_2 h \cos\theta)^{3/2}} + \rho_{22} \int_{\alpha_2}^{\pi-\alpha_2} \frac{\sin^3\theta \, d\theta}{(R_2^2 + h^2 - 2R_2 h \cos\theta)^{3/2}} + \rho_{21} \int_{\pi-\alpha_2}^{\pi} \frac{\sin^3\theta \, d\theta}{(R_2^2 + h^2 - 2R_2 h \cos\theta)^{3/2}} \right\}$$

where
$$\rho_{22} = -\frac{1-\cos\alpha_2}{\cos\alpha_2} \rho_{21}$$

With $\mu_0 = 4\pi 10^{-7}$ Hm$^{-1}$, the parameter $\rho_{ij}$ represents the mean values of the electric charges (in coulombs/m$^2$) of the inner core ($i=1$) and of the outer core periphery ($i=2$) near the poles ($j=1$) and at the equator ($j=2$). $\Omega_1$ and $\Omega_2$ are respectively the angular velocities of the inner core and of the mantle, and $\theta$ and $\alpha_i$ are centre angles to the northwards rotational axis ; $\alpha_i$ separates the areas of positive and negative charges (see Fig. 5 where the charges are schematically represented for the present time) : $i=1$ refers to the inner core and $i=2$ to the periphery of the outer core. $R_1=1270*10^3$ metres (inner core radius), $R_2=3470*10^3$ metres (outer core radius), $h=6370*10^3$ metres (the Earth's radius).

We must evaluate $\Omega_1$ and $\rho_{i1}$. The present angular velocity of the mantle $\Omega_2$ is $72.72*10^{-6}$ radians/second. We suppose that $\Omega_1$ is constant with time and that $Bz=0$ when $\Omega_1 = \Omega_2$. We can assess this value of $\Omega_2$ from the present observation that day time increases by 0.0016 second per century. If we put this effect down to the slowing down of the mantle, in conformity with our theory, we can consider this phenomenon as identical to the one which ought to have occurred between 205 and 65 Millions Years before Present (see Fig.1). That means that, in 140 Millions Years, there will be 2240 seconds (a little more than 37 minutes) which will have been added to the present 24 hours for one mantle rotation. The mantle angular velocity $\Omega_2$ will be then the same as that of the inner core angular velocity $\Omega_1$, that is to say $70.88*10^{-6}$ radians per second.

As for the values of $\rho_{ij}$, the constraints are the number of atoms of iron ($1.078*10^{22}$/g), the presence of some per cents of nickel being not able to consistently modify this number. The maximum electrical charges for $Fe^{++}$ is therefore 6.2 coulombs/m² and for $Fe^{+++}$ 9.27 coulombs/m². We have chosen the intermediate value of 8 coulombs/m². For the present value of $Bz=-0.5*10^{-4}$ Tesla (see Lorrain et al. (2000) about the sign), if $\rho_{11}=-8$ cb/m² and $\rho_{21}=+24$ cb/m², $\alpha_1=1$ rad (57.3°) and $\alpha_2=0.7$ rad (40.1°). The electrical charges are negative at the poles of the inner core but must be positive at the poles of the outer core. The hypothesis is that the heat produced at the equatorial region by the friction of the inner core generates warm streams inside the liquid outer core, which reach the poles of the outer core because of convection and Coriolis forces, and are sufficient to reverse the heat gradients at the outer core periphery. The value of $\rho_{21}$ is three times higher than the chosen electrical charges for $Fe^{+++}$. There are probably several individualized conductive layers in the periphery of the outer core. Besides, we consider that $\rho_{11}$ remains steady because of the thermal inertia of the solid inner core, but $\rho_{21}$ must vary (linearly ?) with respect to the absolute difference of the angular velocities of the mantle and inner core. Its value when $\Omega_1=\Omega_2$ would be 0.7 cb/m² with the same other parameters, which can be considered as a kind of thermo electric remanence.

The warm streams are possible because the mantle rotates faster than the inner core ; they are probably disturbed and blocked up when the mantle rotates less rapidly than the inner core. Thus the heat gradients are decreasing from the equators to the poles for both the inner core and the periphery of the outer core, $\rho_{11}$ and $\rho_{21}$ are then both negative, which involves a reversed geomagnetic field.

This effect occurs because $|Bz_1|$ is fairly small and the sign of $Bz$ is rapidly given by $Bz_2$ as $Bz_2$ is increasing. For example, for the current value of $Bz$, $|Bz_2|$ is worth 35 times $|Bz_1|$. As a consequence, the dipolar field must be connected to $Bz_2$ (at the periphery of the outer core) and the rapid fluctuations of the geomagnetic field probably reflect a strong instability in time at the boundary between mantle plates and liquid outer core.

This model can explain the numerous magnetic reversals in the early quaternary, at a moment when Bz was close to zero and $\Omega_2 \approx \Omega_1$ (see Fig. 1), by the instability of the warm streams originated at the equator of the inner core, and which could reach the periphery of the outer core (then $Bz < 0$), or could not do it because of insufficient heat (then $Bz > 0$). Besides, $Bz_2$ is very sensitive to the parameter $\alpha_2$, that is to say to the surface of the periphery of the outer core reached by the warm streams. So, from the current values, Bz would decrease by 1/10 if $\alpha_2$ increased of 0.14 radian (8°), that is to say following a small extension of the warm polar area of the periphery of the outer core. Even if these streams are well stabilized, some weak variations of $\alpha_2$ can still involve the observed present rapid variations of the geomagnetic field magnitudes. As a consequence of the link between the mean temperatures inside the troposphere and the geomagnetic field magnitudes, some fairly weak disturbances in the warm streams inside the outer core (causing variations of $\alpha_2$) occurring at the time of high absolute value of the geomagnetic field magnitudes, as should have been the case one million years ago, must have caused noteworthy variations of Bz. Consequently the well known glacial-interglacial stage alternations occurred at this time.

**Conclusion**

The source of the Earth's activity lies in the difference of the angular velocities of the mantle and of the solid inner core. The friction between both spheres generates heat, which is the cause of the melted iron constituting mainly the liquid outer core, as well as the source of the global heat flow. The solid inner core angular velocity is supposed to remain steady, while the mantle angular velocity depends on gyroscopic forces (involving acceleration) and slowing down due to external attractions and mainly to the motions of mantle plates 2900 km thick.

The difference of tangential velocity between the equator and the poles at the friction locations creates differences of heat , which are the source of thermo electric effects inside conductive elements.

So the geomagnetic model of two electrically charged shells, located on the surface of the solid inner core and at the periphery of the liquid outer core permits us to represent the geomagnetic field, with its dipole and non dipole components, as well as its past variations, including the reverse field.

The variations of the geomagnetic field are therefore direct consequence of the variations of the angular velocity of the mantle relative to that of the inner core. In one word, the biological and tectonic evolutions during the geological times are due to those phenomena. So, the limits of eras coincide exactly with the passage to zero of the geomagnetic field.

This model has three advantages : (1) it fits correctly with all the known geophysical and geological elements, eliminating the contradictory explanations such as the convection currents in solid materials or the homogenous heat flow all over the Globe from the radioactivity inside continental rocks, (2) it is simple and logical from a physical point of view, and (3) it purchases a simple and obvious explanation of the geomagnetic field.

The cycles of about 230-250 million years, which show the correlation between the mantle angular velocity variations, the geomagnetic variations, and therefore the climate, allow us to predict future events. Therefore, the global warming observed since the beginning of the last century (involving glacier melting, for example), is parallel to the Earth's magnetic field decrease, and is going to continue. Stopping atmospheric pollution is necessary, but that will unfortunately not stop this warming which we will have to adapt to. It is the same for the forests which are reputed to be damaged by acid rains, but which may be spoiled in reality from their roots by acid liquid arising inside the cracks of the crust.

Last but not least, this model should enable us to examine the mechanisms which trigger earthquakes from a stress anisotropy point of view. Earthquake prediction could then be made from continuous concrete determinations in the field of the stress anisotropy, knowing the thresholds of rupture.

As an extension to the understanding of other planets of the solar system, the presence or absence of magnetic field, as well as of heat source, can be investigated on the ground of the probability of the disjunction between the angular velocity of their core and of their envelope. For example, the magnetic field of Uranus, which is tilted about 60 degrees with respect to the rotation axis, and as a result, moves like a corkscrew as Uranus rotates, may be the result of two electrically charged shells, the one "stuck" to the envelope rotating around its axis lying 8 degrees out of the plane of the orbit, the other "stuck" to the core rotating around an axis probably closer to the perpendicular of the plane of orbit.

**Acknowledgments**

I am very grateful to J.C. Gianduzzo, and J. Salardenne (Université Bordeaux 1) for their work and advice in electro-magnetism, J.C. Batsale (Université Bordeaux 1) for his advice in thermal diffusion, P. Dordor (Université Bordeaux 1) for his advice in thermoelectricity, D. Fein (University of San Diego) and E. Strouse (Université Bordeaux 1) for their suggestions and corrections.

**Figures**

Figure 1 : Diagram showing the tight connection between geomagnetic field, mantle angular velocity, and climates in the geological times.

Figure 2 : Scheme showing the possible cracks originating earthquakes in the case of strike-slip faults.

Figure 3 : Block diagrams representing the upward magmatic flow supposed to occur beneath the Mediterranean region *(Rousseau, 1987 ; Rousseau, 1992)*.

Figure 4 : Scheme showing the consequences of the mechanism of Fig. 3 on the sharing of the constraints inside the Mediterranean lithosphere *(Rousseau, 1987 ; Rousseau, 1992)*.

> A= only extension occurs under the effect of the magma pressure (heavy arrow) ; B and b = zones where no deformation occurs at the top and the bottom of the whole plate ; C = zone where extension and compression occur under the effect of the flexion ; a = the opposite, in direction, of A : the loading is here the dead weight represented by the double arrow.

Figure 5 : Schematic representation of the present inner dynamics of the Earth.

> The + and - indicate the electrically charged areas. On this scaled sketch, the crust is not indicated because of its very small thickness.

Figure 1 : Diagram showing the tight connection between geomagnetic field, mantle angular velocity, and climates in the geological times.

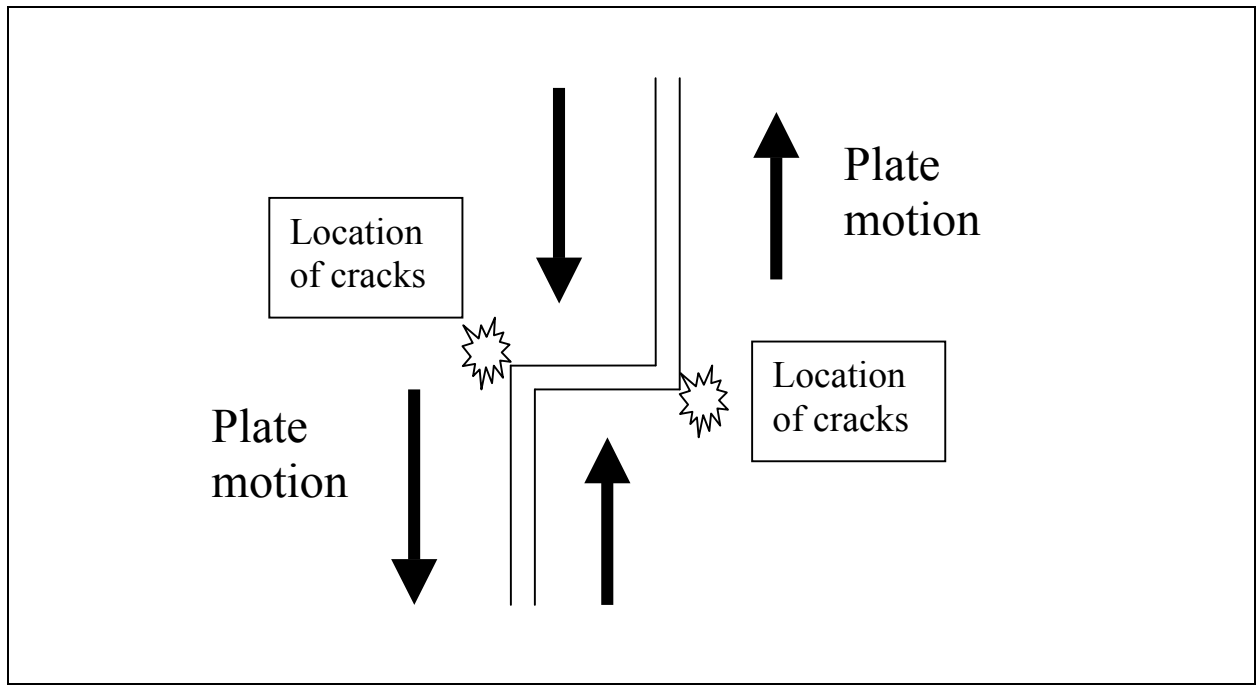

Figure 2 : Scheme showing the possible cracks originating earthquakes in the case of strike-slip faults.

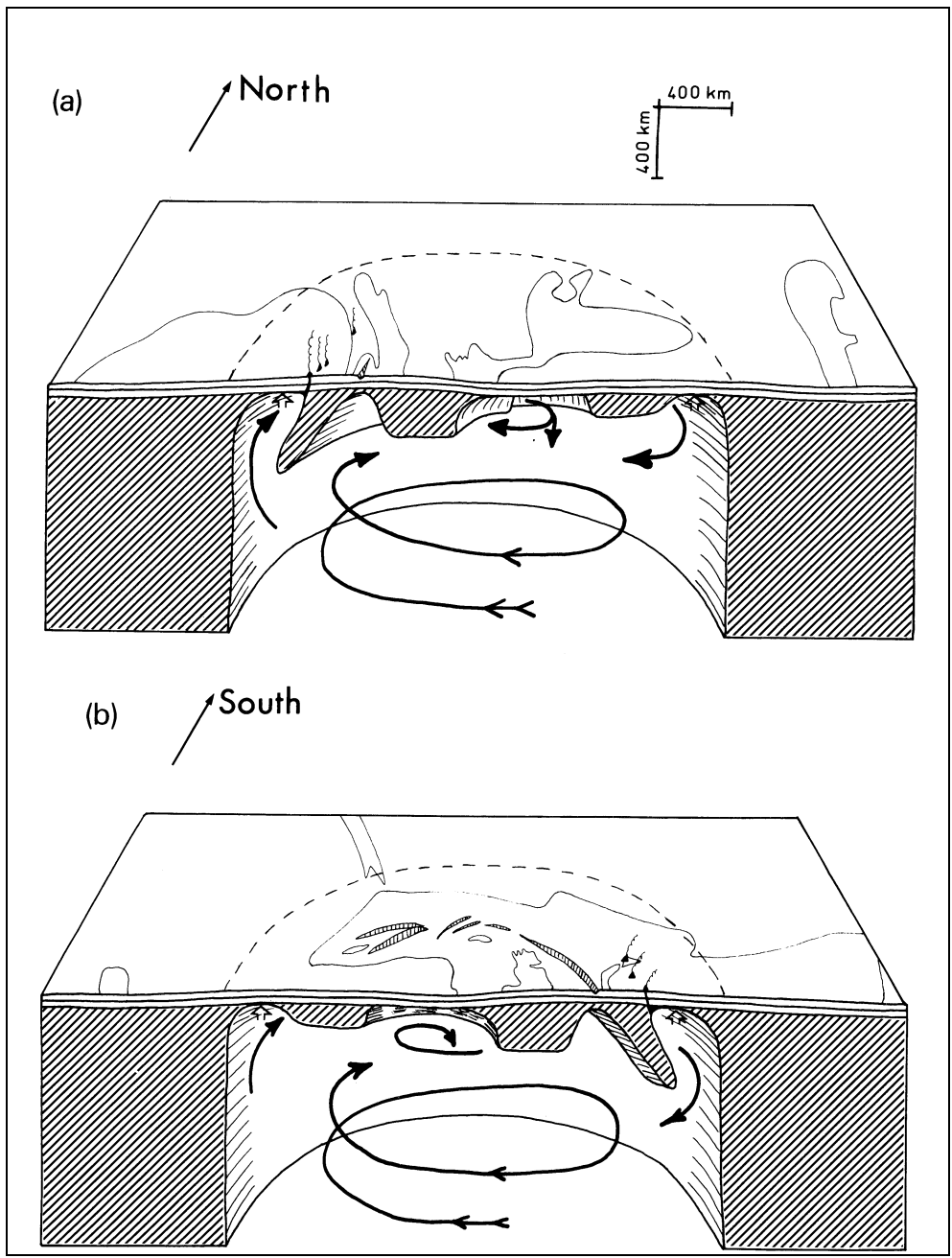

Figure 3 : Block diagrams representing the upward magmatic flow supposed to occur beneath the Mediterranean region *(Rousseau, 1987 ; Rousseau, 1992)*.

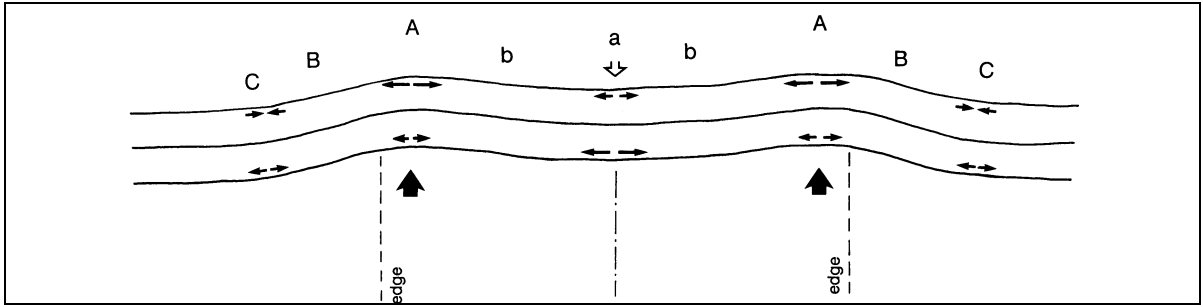

Figure 4 : Scheme showing the consequences of the mechanism of Fig. 3 on the sharing of the constraints inside the Mediterranean lithosphere *(Rousseau, 1987 ; Rousseau, 1992)*.
A= only extension occurs under the effect of the magma pressure (heavy arrow) ; B and b = zones where no deformation occurs at the top and the bottom of the whole plate ; C = zone where extension and compression occur under the effect of the flexion ; a = the opposite, in direction, of A : the loading is here the dead weight represented by the double arrow.

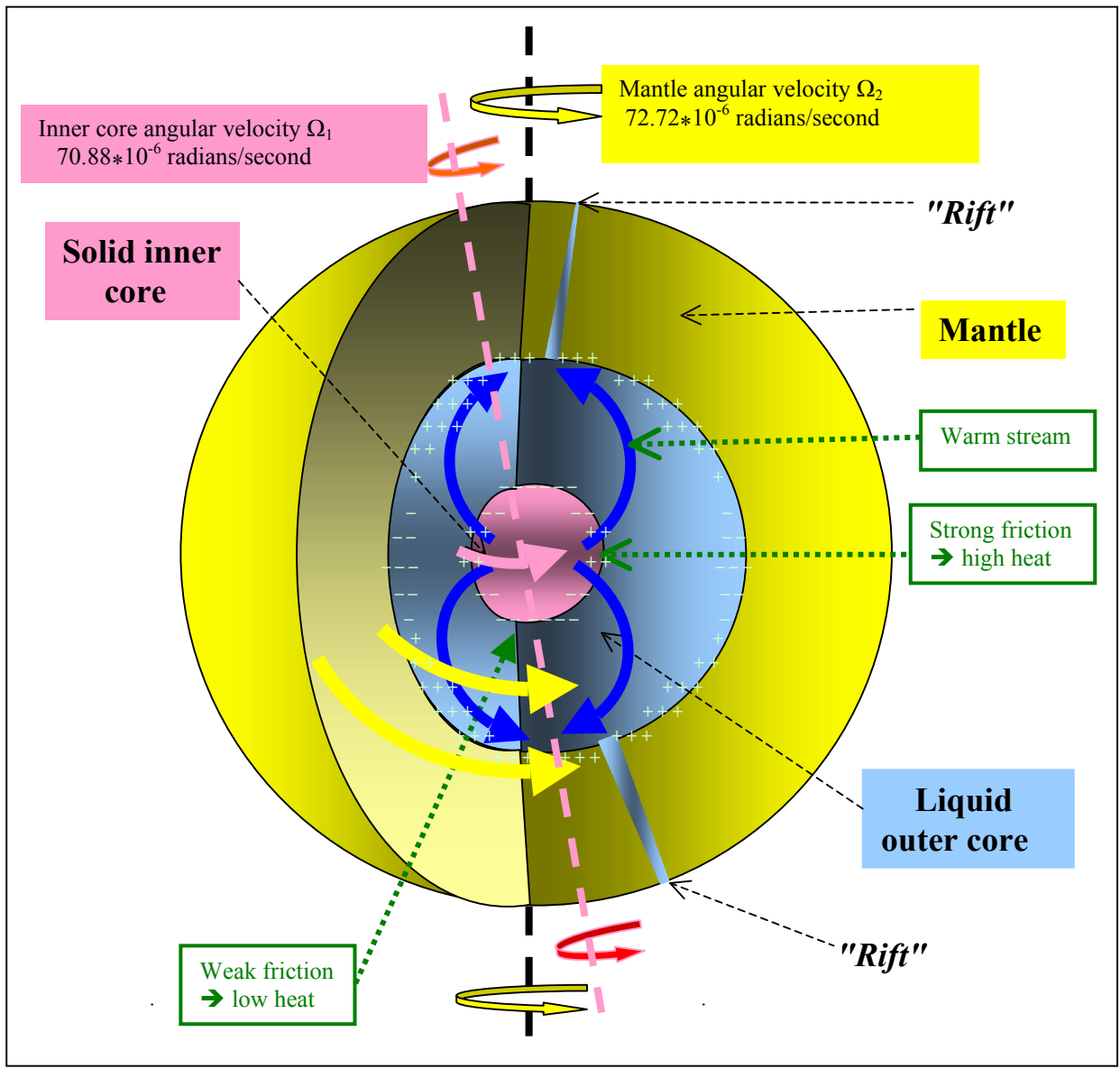

Figure 5 : Schematic representation of the present inner dynamics of the Earth.
The + and - indicate the electrically charged areas. On this scaled sketch, the crust is not indicated because of its very small thickness.